\documentstyle[aps,prl,epsf]{revtex}

\tightenlines

\begin{document}

\twocolumn[\hsize\textwidth\columnwidth\hsize\csname
  @twocolumnfalse\endcsname
\draft

\title{Violation of Time Reversal Invariance\\ in the Decays
$K_L \rightarrow \pi^+ \pi^- \gamma$ and $K_L \rightarrow \pi^+ \pi^- e^+ e^-$}

\author{L.~M.~Sehgal and J.~van~Leusen}
\address{Institute of Theoretical Physics, RWTH Aachen, D-52056 Aachen, Germany}

\maketitle

\begin{abstract}
The origin of the large $CP$-odd and $T$-odd asymmetry observed in the decay 
$K_L \rightarrow \pi^+ \pi^- e^+ e^-$ is traced to the polarization properties of 
the photon in the decay $K_L \rightarrow \pi^+ \pi^- \gamma$. The Stokes vector
of the photon $\vec{S} = \left( S_1,\, S_2,\, S_3\right)$ is studied as a function 
of the photon energy and found to possess $CP$-violating components $S_1$ and 
$S_2$ which are sizeable over a large part of the phase space, despite being 
proportional to the $\epsilon$ parameter of the $K_L$ wave function. The 
component $S_2$ is $T$-even and manifests itself as a circular polarization of 
the photon, while $S_1$ is $T$-odd and gives rise to the asymmetry observed in
$K_L \rightarrow \pi^+ \pi^- e^+ e^-$. The latter is shown to survive in the 
``hermitian'' limit in which all unitarity phases are absent, and represents a 
genuine example of time reversal symmetry breaking in a $CPT$ invariant theory.  
\end{abstract}
] 

The KTeV experiment has reported the observation of a large $CP$-violating, 
$T$-odd asymmetry in the decay $K_L \rightarrow \pi^+ \pi^- e^+ e^-$~\cite{KTeVcoll},
in agreement with a theoretical prediction made some years 
ago~\cite{Sehgal:Wanninger,Heiliger:Sehgal}. In this letter, we trace the origin 
of this effect to a large violation of $CP$-invariance and $T$-invariance in the 
decay $K_L \rightarrow \pi^+ \pi^- \gamma$, which is hidden in the polarization 
state of the photon. We explain why the effect is large, despite the fact that it 
stems entirely from the $\epsilon$-impurity of the $K_L$ wave function. Our 
analysis demonstrates that the $T$-odd asymmetry does not vanish in the limit in 
which unitarity phases, expressing the non-hermiticity of the effective Hamiltonian,
are switched off, and thus represents a genuine example of time reversal 
non-invariance.

The decay $K_L \rightarrow \pi^+ \pi^- \gamma$ is known empirically~\cite{E731coll} 
to contain a bremsstrahlung component ($I\!B$) as well as a direct emisson component
($D\!E$), with a relative strength $D\!E / (D\!E + I\!B) = 0.68$ for photons above 
$20 \, MeV$. By contrast, the decay $K_S \rightarrow \pi^+ \pi^- \gamma$ is well 
reproduced by pure bremsstrahlung. The simplest matrix element consistent with these 
features is~\cite{Sehgal:Wanninger}
\begin{eqnarray}
{\cal M} (K_S \rightarrow \pi^+ \pi^- \gamma) & = & e f_S
\left[\frac{\epsilon \cdot p_+}{k \cdot p_+} -
\frac{\epsilon \cdot p_-}{k \cdot p_-} \right] \nonumber\\
{\cal M} (K_L \rightarrow \pi^+ \pi^- \gamma) & = & e f_L
\left[\frac{\epsilon \cdot p_+}{k \cdot p_+} -
\frac{\epsilon \cdot p_-}{k \cdot p_-} \right]
\label{matelem}\\
& & \mbox{} + e \frac{f_{DE}}{{M_K}^4}
\epsilon_{\mu\nu\rho\sigma} \epsilon^{\mu}k^{\nu}{p_+}^{\rho}{p_-}^{\sigma}
\nonumber
\end{eqnarray}
where
\begin{eqnarray}
f_L \equiv |f_S| g_{Br},\: g_{Br} = \eta_{+-} e^{i \delta_0(s={M_K}^2)},
\nonumber\\
f_{DE} \equiv |f_S| g_{M1},\: g_{M1} = i(0.76)e^{i \delta_1(s)}.
\label{whatmeansf}
\end{eqnarray}
Here the direct emission has been represented by a $CP$-conserving magnetic
dipole coupling $g_{M1}$, whose magnitude $|g_{M1}| = 0.76$ is fixed by the
empirical ratio $D\!E/I\!B$. The phase factors appearing in $g_{Br}$ and $g_{M1}$
are dictated by the Low theorem for bremsstrahlung,
and the Watson theorem for final state interactions. The factor $i$ in
$g_{M1}$ is a consequence of $CPT$ invariance~\cite{Sehgal:Costa:Lee}.
The matrix element for $K_L \rightarrow \pi^+ \pi^- \gamma$ contains 
simultaneously electric multipoles associated with
bremsstrahlung ($E1,\,E3,\,E5, \, \cdots$), which have $CP=+1$, 
and a magnetic $M1$
multipole with $CP = -1$. It follows that interference of the electric and
magnetic emissions should give rise to $CP$-violation.

To determine the nature of this interference, we write the
$K_L \rightarrow \pi^+ \pi^- \gamma$ amplitude more generally as
\begin{eqnarray}
{\cal M}(K_L \rightarrow \pi^+ \pi^- \gamma) & = &
\frac{1}{{M_K}^3} \left\{ E(\omega,\cos \theta) \right. \\
& & \left. \times \left[\epsilon \cdot p_+ \, k
\cdot p_- - \epsilon \cdot p_- \, k \cdot p_+ \right] \right. \nonumber\\
& & \left. \mbox{} + M(\omega, \cos \theta)
\epsilon_{\mu\nu\rho\sigma}\epsilon^{\mu}k^{\nu}{p_+}^{\rho}{p_-}^{\sigma}
\right\} \nonumber
\end{eqnarray}
where $\omega$ is the photon energy in the $K_L$ rest frame, and $\theta$
is the angle between $\pi^+$ and $\gamma$ in the $\pi^+ \pi^-$ rest frame.
In the model represented by Eqs. (\ref{matelem}) and (\ref{whatmeansf}),
the electric and magnetic amplitudes are (omitting a common factor 
$e |f_S| / M_K$)
\begin{eqnarray}
E & = & \left( \frac{2M_K}{\omega} \right)^2 \frac{g_{Br}}{1-\beta^2 \cos^2 \theta}
\nonumber\\
M & = & g_{M1}
\label{defEandM}
\end{eqnarray}
where $\beta = (1- 4 {m_{\pi}}^2/s)^{1/2}$, $\sqrt{s}$ being the $\pi^+\pi^-$ invariant
mass. The Dalitz plot density, summed over photon polarizations is
\begin{eqnarray}
\frac{d \Gamma}{d \omega \, d\! \cos \theta} & = & \frac{1}{512 \pi^3}
\left( \frac{\omega}{M_K} \right)^3 \beta^3 \left( 1- \frac{2 \omega}{M_K}
\right) \nonumber\\
& & \times \sin^2 \theta \left[ |E|^2 + |M|^2 \right].
\label{dGammaint}
\end{eqnarray}
Clearly, there is no interference between the electric and magnetic multipoles
if the photon polarization is unobserved. Therefore, any $CP$-violation
involving the interference of $g_{Br}$ and $g_{M1}$ is encoded in the
polarization state of the photon. 

The photon polarization can be defined in terms of the density matrix
\begin{equation}
\rho = \left( \begin{array}{cc} |E|^2 & E^*M \\ EM^* & |M|^2 \end{array} 
\right) = \frac{1}{2} \left( |E|^2+|M|^2 \right) \left[ 1 \!\!\!\!\:\: \mbox{l}
 + \vec{S} \cdot \vec{\tau}  \right]
\end{equation}
where $\vec{\tau} = (\tau_1, \, \tau_2, \, \tau_3)$ denotes the Pauli matrices,
and $\vec{S}$ is the Stokes vector of the photon with components
\begin{eqnarray}
S_1 & = & 2 Re \left( E^*M \right) / \left( |E|^2 + |M|^2 \right) \nonumber\\
S_2 & = & 2 Im \left( E^*M \right) / \left( |E|^2 + |M|^2 \right) \\
S_3 & = & \left(|E|^2 - |M|^2 \right) / \left( |E|^2 + |M|^2 \right). \nonumber
\end{eqnarray}
The component $S_3$ measures the relative strength of the electric and magnetic
radiation at a given point in the Dalitz plot. The effects of $CP$-violation
reside in the components $S_1$ and $S_2$, which are proportional to
$Re \, ({g_{Br}}^* g_{M1})$ and $Im \, ({g_{Br}}^* g_{M1})$, respectively.
Physically, $S_2$ is the net circular polarization of the photon: it is proportional 
to the difference of $\left| E - iM \right|^2$ and $\left| E + iM \right|^2$, 
which are the probabilities for left-handed and right-handed radiation. Such a 
polarization is a $CP$-odd, $T$-even effect, which is known to be possible 
in decays like $K_L \rightarrow \pi^+ \pi^- \gamma$ or
$K_{L,S} \rightarrow \gamma \gamma$ whenever there is $CP$-violation accompanied
by unitarity phases~\cite{Sehgal:Costa:Lee,Sehgal}. 
To understand the significance of $S_1$,
we examine the dependence of the $K_L \rightarrow \pi^+ \pi^- \gamma$ decay
on the angle $\phi$ between the polarization vector $\vec{\epsilon}$ and
the unit vector $\vec{n}_{\pi}$ normal to the decay plane (we
choose coordinates such that $\vec{k} = (0,\, 0,\, k)$, $\vec{n}_{\pi} =
(1,\, 0, \, 0)$, $\vec{p}_+ = (0,\, p\sin \theta, \, p\cos \theta)$ and
$\vec{\epsilon} = (\cos \phi, \, \sin \phi, \, 0)$):
\begin{eqnarray}
\frac{d \Gamma}{d \omega \, d\! \cos \theta \, d \phi} & \sim &
\left| E \sin \phi - M \cos \phi \right|^2 \nonumber\\
& \sim & 1 - \left[ S_3 \cos 2\phi + S_1 \sin 2\phi \right].
\label{distS}
\end{eqnarray}
Notice that the Stokes parameter $S_1$ appears as a coefficient of a term 
$\sin 2 \phi$ which changes sign under $CP$ as well as $T$. Thus $S_1$ is a
measure of a $CP$-odd, $T$-odd correlation. 
The essential idea of Refs.~\cite{Sehgal:Wanninger,Heiliger:Sehgal}
is to use in place of $\vec{\epsilon}$,
the vector $\vec{n}_l$ normal to the plane of the Dalitz pair in the reaction
$K_L \rightarrow \pi^+ \pi^- \gamma^* \rightarrow \pi^+ \pi^- e^+ e^-$. This
motivates the study of the distribution $d\Gamma/d\phi$ in the decay
$K_L \rightarrow \pi^+ \pi^- e^+ e^-$, where $\phi$ is the angle between
the $\pi^+ \pi^-$ and $e^+ e^-$ planes.

\begin{figure}
\centerline{\epsfxsize=8.0cm \epsfysize=7.3cm \epsffile{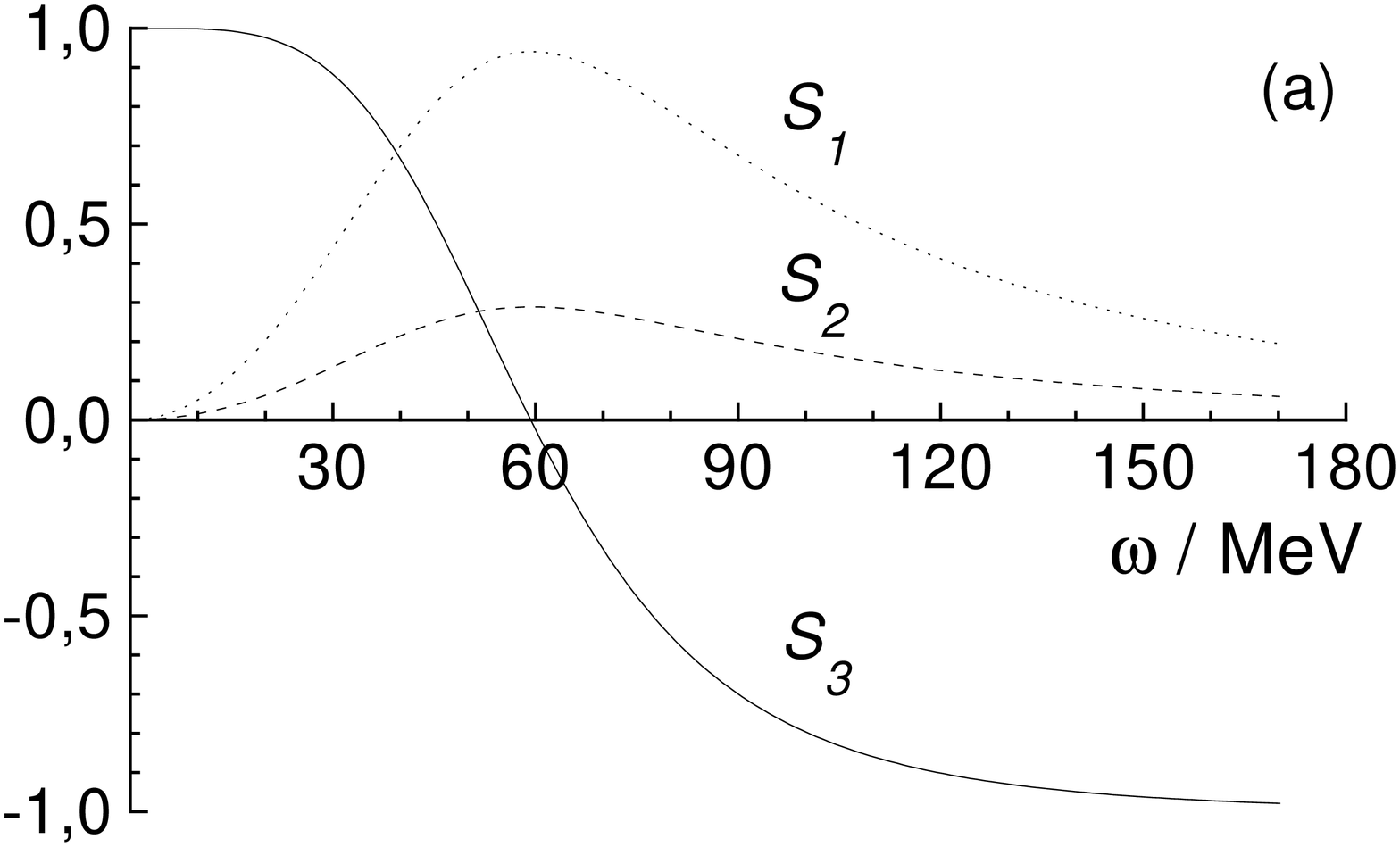}}
\centerline{\epsfxsize=8.0cm \epsfysize=7.3cm \epsffile{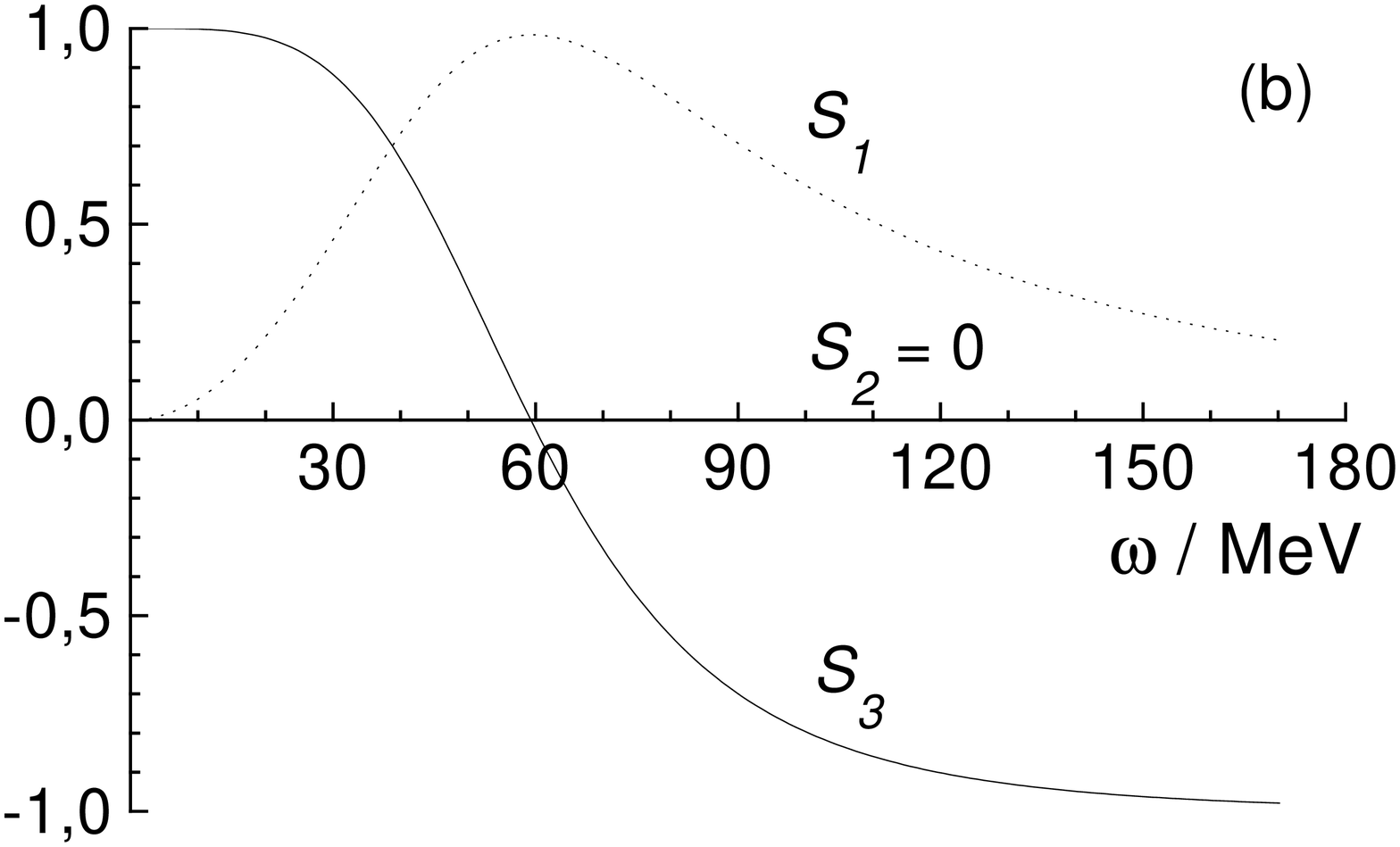}}
\centerline{\epsfxsize=8.0cm \epsfysize=7.3cm \epsffile{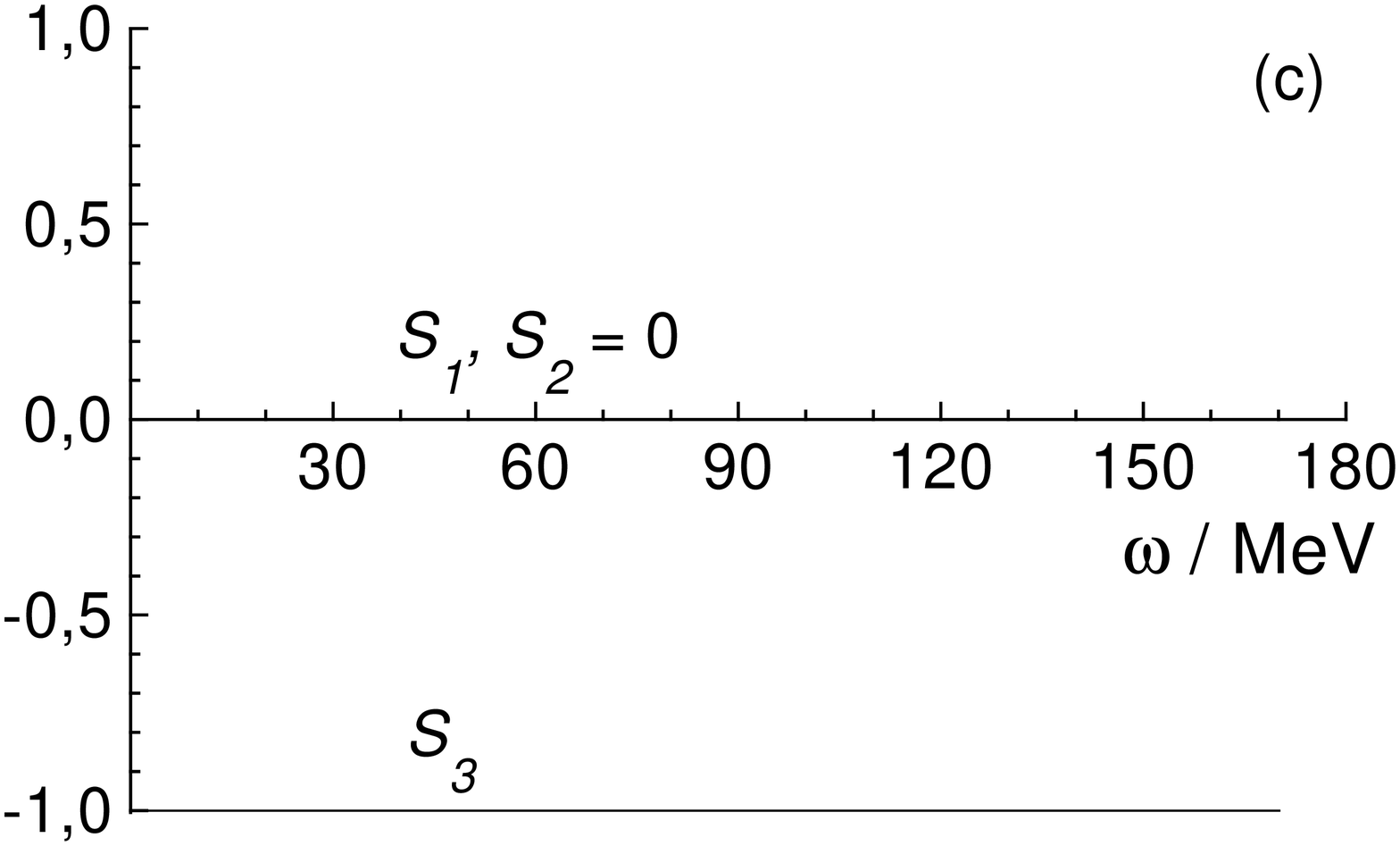}}
\caption{(a) Stokes parameters of photon in
$K_L \rightarrow \pi^+ \pi^- \gamma$; (b) Hermitian limit
$\delta_0 = \delta_1 = 0$, $ar\!g \, \epsilon = \pi / 2$;
(c) $CP$-invariant limit $\epsilon \rightarrow 0$.
\label{SivsE}
}
\end{figure}

To obtain a quantitative idea of the magnitude of $CP$- violation in
$K_L \rightarrow \pi^+ \pi^- \gamma$, we show in Fig.~\ref{SivsE}a the three
components of the Stokes  vector as a function of the photon
energy. These are calculated from the amplitudes
(\ref{defEandM}) using weighted averages of $|E|^2$, $|M|^2$, $E^*M$
and $EM^*$ over $\cos \theta$~\cite{fussn}. The values of $S_1$ and $S_2$
are remarkably large, considering that the only assumed source of $CP$-violation 
is the $\epsilon$-impurity in the $K_L$ wave-function ($\epsilon = \eta_{+-}$).
Clearly the factor $(2 M_K/\omega)^2$  in $E$ enhances it to a level
that makes it comparable to the $CP$-conserving amplitude $M$. This is
evident from the behaviour of the parameter $S_3$, which swings from a
dominant electric behaviour at low $\omega$ ($S_3 \approx 1$) to a
dominant magnetic behaviour at large $\omega$ ($S_3 \approx -1$),
with a zero in the region $\omega \approx 60 \, MeV$. The essential 
difference between the $T$-odd parameter $S_1$ and the $T$-even parameter
$S_2$ comes to light when we compare their behaviour
in the ``hermitian'' limit: this is the limit in which the $T$-matrix or
effective Hamiltonian governing the decay $K_L \rightarrow \pi^+ \pi^- \gamma$
is taken to be hermitian, all unitarity phases related to real intermediate
states being dropped. This limit is realized by taking
$\delta_0 , \, \delta_1 \rightarrow 0$, and $ar\!g \, \epsilon \rightarrow \pi /2$.
The last of these follows from the fact that $\epsilon$ may be written as
\begin{equation}
\epsilon = \frac{\Gamma_{12}-\Gamma_{21} + i \left( M_{12}-M_{21} \right)}
{\gamma_S - \gamma_L - 2 i \left( m_L - m_S \right)}
\end{equation}
where $H_{eff} = M - i\Gamma$ is the mass matrix of the $K^0$-$\overline{K}^0$
system. The hermitian limit obtains when $\Gamma_{12} = \Gamma_{21} = \gamma_S
= \gamma_L = 0$. As seen from Fig.~\ref{SivsE}b, $S_2$ vanishes in this
limit, but $S_1$ survives, as befits a $CP$-odd, $T$-odd observable. This
difference in behaviour is obvious from the fact that in the hermitian limit
\begin{eqnarray}
S_1 & \sim & Re({g_{Br}}^* g_{M1}) \sim \sin(\phi_{+-} + \delta_0 - \delta_1)
\rightarrow 1 \nonumber\\
S_2 & \sim & Im({g_{Br}}^* g_{M1}) \sim \cos(\phi_{+-} + \delta_0 - \delta_1)
\rightarrow 0
\end{eqnarray}
Fig.~\ref{SivsE}c shows what happens in the $CP$-invariant limit 
$\epsilon \rightarrow 0$: the parameters $S_1$, $S_2$ collapse to zero, while 
$S_3$ attains the uniform value $-1$.
It is clear that we are dealing here with an exceptional
situation in which a $CP$-impurity of a few parts in a thousand
in the $K_L$ wave-function is magnified into a huge $CP$-odd, $T$-odd effect
in the photon polarization.

We can now examine how these large $CP$-violating effects are transported
to the decay $K_L \rightarrow \pi^+ \pi^- e^+ e^-$. The matrix element for
$K_L \rightarrow \pi^+ \pi^- e^+ e^-$ can be written as~\cite{Sehgal:Wanninger,Heiliger:Sehgal}
\begin{eqnarray}
{\cal M} (K_L \rightarrow \pi^+ \pi^- e^+ e^-) & = & {\cal M}_{br} +
{\cal M}_{mag} \nonumber\\
& & \mbox{} + {\cal M}_{CR} + {\cal M}_{SD}.
\end{eqnarray}
Here ${\cal M}_{br}$ and ${\cal M}_{mag}$ are the conversion amplitudes
associated with the bremsstrahlung and $M1$ parts of the
$K_L \rightarrow \pi^+ \pi^- \gamma$ amplitude. In addition, we have
introduced an amplitude ${\cal M}_{CR}$ denoting $\pi^+ \pi^-$ production
in a $J = 0$ state (not possible in a real radiative decay), as well as an
amplitude ${\cal M}_{SD}$ associated with the short-distance interaction
$s \rightarrow d\, e^+ e^-$. The last
of these turns out to be numerically negligible because of the smallness
of the $C\!K\!M$ factor $V_{ts} {V_{td}}^*$. The $s$-wave amplitude
${\cal M}_{CR}$, if approximated by the $K^0$ charge radius diagram,
makes a small ($\sim 1 \%$) contribution to the decay rate. Thus the
dominant features of the decay are due to the conversion amplitude
${\cal M}_{br} + {\cal M}_{mag}$.

Within such a model, one can calculate the differential decay rate in the
form~\cite{Heiliger:Sehgal}
\begin{equation}
d \Gamma = I(s_{\pi}, \, s_l, \, \cos \theta_l, \, \cos \theta_{\pi}, \, \phi)
\, ds_{\pi} \, ds_l \, d\! \cos \theta_l \, d\! \cos \theta_{\pi} \, d\phi.
\end{equation}
Here $s_{\pi}$ ($s_l$) is the invariant mass of the pion (lepton) pair,
and $\theta_{\pi}$ ($\theta_l$) is the angle of the $\pi^+$ ($l^+$) in the
$\pi^+ \pi^-$ ($l^+ l^-$) rest frame, relative to the dilepton (dipion)
momentum vector in that frame. The all-important variable $\phi$ is defined
in terms of unit vectors constructed from the pion momenta $\vec{p_{\pm}}$
and lepton momenta $\vec{k_{\pm}}$ in the $K_L$ rest frame:
\begin{eqnarray}
\vec{n}_{\pi} & = & \left( \vec{p}_+ \times \vec{p}_- \right) /
\left| \vec{p}_+ \times \vec{p}_- \right|, \nonumber\\
\vec{n}_l & = & \left( \vec{k}_+ \times \vec{k}_- \right) /
\left| \vec{k}_+ \times \vec{k}_- \right|, \nonumber\\
\vec{z} & = & \left( \vec{p}_+ + \vec{p}_- \right) /
\left| \vec{p}_+ + \vec{p}_- \right|, \nonumber
\end{eqnarray}
\begin{eqnarray}
\sin \phi \: = & \vec{n}_{\pi} \times \vec{n}_l \cdot \vec{z} & \:\: (CP = -, T = -),
\nonumber\\
\cos \phi \: = & \vec{n}_{\pi} \cdot \vec{n}_l & \:\:(CP = +, T = +).
\end{eqnarray}
In Ref.~\cite{Sehgal:Wanninger}, an analytic expression was derived for
the 3-dimensional distribution $d\Gamma/ ds_l \, ds_{\pi} \, d\phi$, which
has been used in the Monte Carlo simulation of this
decay. In Ref.~\cite{Heiliger:Sehgal}, a formalism was presented for
obtaining the fully differential decay function $I(s_{\pi},\, s_l, \, \cos
\theta_l, \, \cos \theta_{\pi}, \phi)$.

The principal results of the theoretical model discussed
in~\cite{Sehgal:Wanninger,Heiliger:Sehgal} are as follows:

1. Branching ratio: This was calculated to be~\cite{Sehgal:Wanninger}
\begin{eqnarray}
B\!R(K_L \rightarrow \pi^+ \pi^- e^+ e^-) & = & (1.3 \times 10^{-7})_{Br}
\nonumber\\
& & \mbox{} + (1.8 \times 10^{-7})_{M1} \nonumber \\
& & \mbox{} + (0.04 \times 10^{-7})_{CR} \nonumber\\
& \approx  & 3.1 \times 10^{-7},
\end{eqnarray}
which agrees well with the result $(3.32 \pm 0.14 \pm 0.28) \times 10^{-7}$ 
measured in the
KTeV experiment~\cite{KTeVcoll}. (A preliminary branching ratio 
$2.9 \times 10^{-7}$ has been reported by NA48~\cite{NA48coll}).

2. Asymmetry in $\phi$ distribution: The model predicts a distribution of
the form
\begin{equation}
\frac{d\Gamma}{d\phi} \sim 1 - \left( \Sigma_3 \cos 2\phi + \Sigma_1 \sin 2\phi
 \right)
\end{equation}
which is in complete analogy with the distribution given by Eq.~(\ref{distS}) 
in the case of $K_L \rightarrow \pi^+ \pi^- \gamma$.
The last term is $CP$- and $T$-violating, and produces an asymmetry
\begin{equation}
{\cal A} = \frac{\left( \int_{0}^{\pi/2} - \int_{\pi/2}^{\pi} +
\int_{\pi}^{3\pi/2} - \int_{3\pi/2}^{2\pi} \right) \frac{d\Gamma}{d\phi} d\phi}
{\left( \int_{0}^{\pi/2} + \int_{\pi/2}^{\pi} +
\int_{\pi}^{3\pi/2} + \int_{3\pi/2}^{2\pi}\right) \frac{d\Gamma}{d\phi} d\phi}
= - \frac{2}{\pi} \Sigma_1.
\end{equation}
The predicted value~\cite{Sehgal:Wanninger,Heiliger:Sehgal} is
\begin{equation}
|{\cal A}| = 15 \% \, \sin (\phi_{+-} + \delta_0({M_K}^2) - \overline{\delta}_1 )
\approx 14 \%
\end{equation}
to be compared with the KTeV result~\cite{KTeVcoll}
\begin{equation}
|{\cal A}|_{KTeV} = (13.6 \pm 2.5 \pm 1.2) \%
\end{equation}
The parameters $\Sigma_3$ and $\Sigma_1$ are calculated to be 
$\Sigma_3 = -0.133$, $\Sigma_1 = 0.23$. The $\phi$-distribution
measured by KTeV agrees with this expectation (after acceptance corrections
made in accordance with the model). It should be noted that the sign of
$\Sigma_1$ (and of the asymmetry ${\cal A}$) depends on whether the
numerical coefficient in $g_{M1}$ is taken to be $+0.76$ or $-0.76$. The
data support the positive sign chosen in Eq. (\ref{whatmeansf}).

\begin{figure}
\centerline{\epsfxsize=8.0cm \epsfysize=7.3cm \epsffile{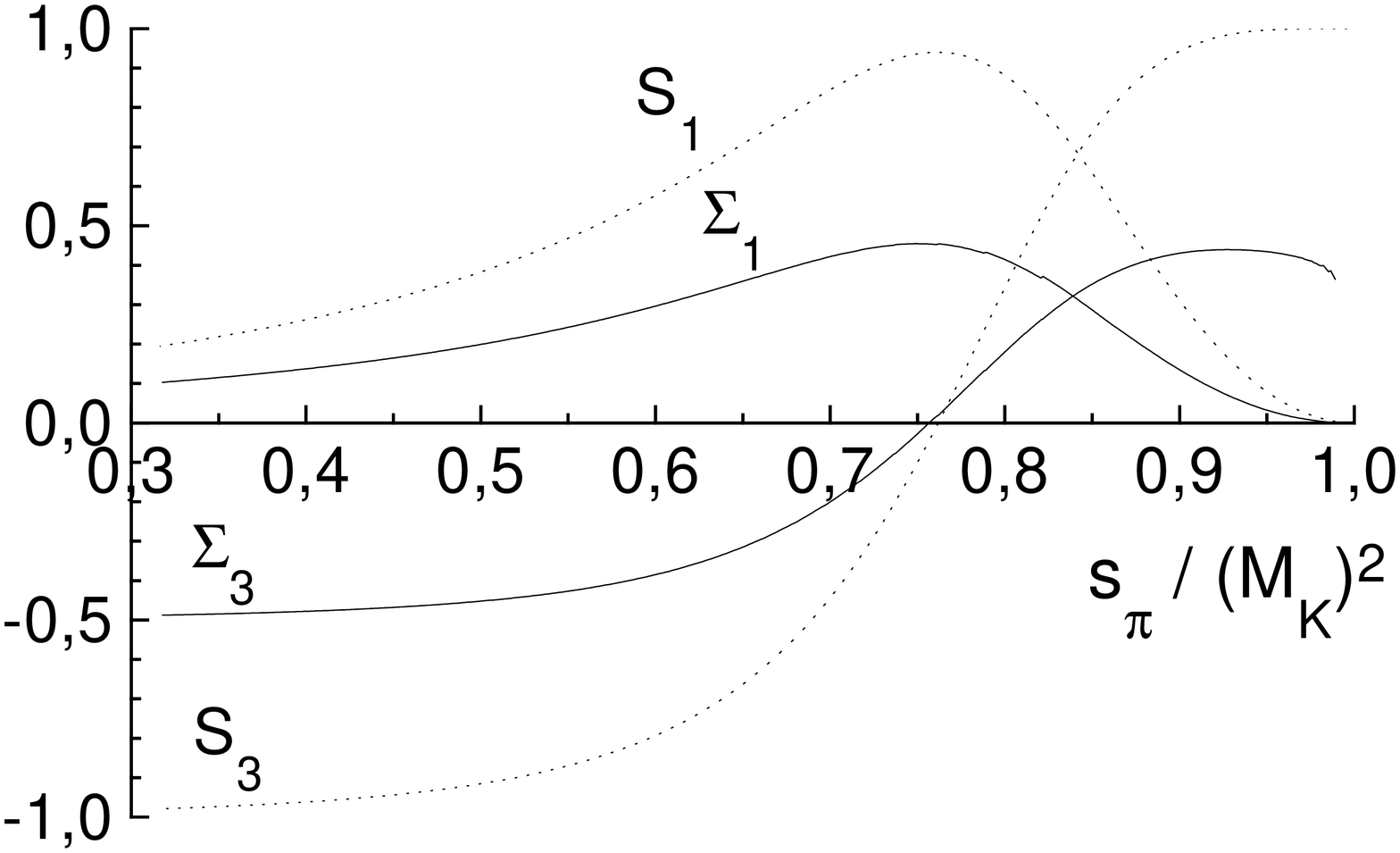}}
\caption{Parameters $\Sigma_1$ and $\Sigma_3$ describing the
$\phi$ - distribution in $K_L \rightarrow \pi^+ \pi^- e^+ e^-$,
compared with the Stokes parameters $S_1$ and $S_3$ in
$K_L \rightarrow \pi^+ \pi^- \gamma$.
\label{SigmavsE}
}
\end{figure}

3. Variation of $\Sigma_{1,3}$ with $s_{\pi}$: As shown in
Fig.~\ref{SigmavsE}, the parameters $\Sigma_1$ and $\Sigma_3$ have a
variation with $s_{\pi}$ that is in close correspondence with the variation
of $S_1$ and $S_3$. (Recall that the photon energy
$\omega$ in $K_L \rightarrow \pi^+ \pi^- \gamma$ can be expressed in terms
of $s_{\pi}$: $s_{\pi} = {M_K}^2 - 2M_K \omega$.) In particular the zero
of $\Sigma_3$ and the zero of $S_3$ occur at almost the same value of $s_{\pi}$.
The similarity in the shape of $\Sigma_1$ and $S_1$ confirms the assertion that 
the asymmetry seen in $K_L \rightarrow \pi^+ \pi^- e^+ e^-$ is related to 
the $CP$-odd, $T$-odd component of the Stokes vector in 
$K_L \rightarrow \pi^+ \pi^- \gamma$. The difference in scale is a measure of 
the analyzing power of the Dalitz pair process, viewed as a probe of the 
photon polarization.\\

Finally, we remark that our analysis takes for granted the validity of $CPT$
invariance in the decays $K_L \rightarrow \pi^+ \pi^- \gamma$ and 
$K_L \rightarrow \pi^+ \pi^- e^+ e^-$. If the assumption of $CPT$ invariance 
is relaxed, the asymmetry observed in the KTeV experiment may be interpreted 
as some combination of $T$- and 
$CPT$-violation~\cite{Wolfenstein:Alvarez:Ellis:Bigi}. From the point of 
view of the present paper, the effect is understandable in a $CPT$-invariant
framework, and follows inexorably from the empirical features of the decays
 $K_{L,S} \rightarrow \pi^+ \pi^- \gamma$ mentioned at the outset.\\

Some of the ideas of this paper were presented by L.~M.~S.\ at the Kaon 99
Conference in Chicago~\cite{Sehgal:Chicago}.

\begin{thebibliography}{99}
\bibitem{fussn} In our numerical work we take $\phi_{+-} = 43^{\circ}$,
$\delta_0 ({M_K}^2) = 40^{\circ}$, and an average value
$\overline{\delta}_1 = 10^{\circ}$.
We have also carried out a calculation which includes the $s$-dependence 
of $\delta_1(s)$, as well as the measured form factor 
$g_{M1}(s)$~\cite{KTeVcoll,E731coll} in place of the constant value
$g_{M1} = 0.76$. The curves in Fig.~\ref{SivsE}a are 
mildly affected, the zero in $S_3$ and the maximum in $S_1$, $S_2$ 
shifting to about $50 \, MeV$.
\end{thebibliography}
\end{document}